\documentclass[journal=nalefd,manuscript=letter]{achemso}

\usepackage[version=3]{mhchem} 
\usepackage{gensymb}
\usepackage{xcolor}

\author{Prince Khatri}
\affiliation[Exeter University]
{College of Engineering, Mathematics and Physical Sciences, University of Exeter, Exeter, EX4 4QF, United Kingdom.}
\author{Andrew J. Ramsay}
\affiliation[Hitachi Cambridge]
{Hitachi Cambridge Laboratory, Hitachi Europe Limited, Cambridge, CB3 0HE, United Kingdom}
\author{Ralph N. E. Malein}
\affiliation[Exeter University]
{College of Engineering, Mathematics and Physical Sciences, University of Exeter, Exeter, EX4 4QF, United Kingdom.}
\author{Harold M. H. Chong}
\affiliation[Southampton]
{Sustainable Electronics Technology Group, School of Electronics and Computer Science, University of Southampton, Southampton, SO17 1BJ, UK}
\author{Isaac J. Luxmoore}
\affiliation[Exeter University]
{College of Engineering, Mathematics and Physical Sciences, University of Exeter, Exeter, EX4 4QF, United Kingdom.}
\email{i.j.luxmoore@exeter.ac.uk}

\title[An \textsf{achemso} demo]
  {Optical control of the charge state of color centers in hexagonal boron nitride}

\begin{document}
\begin{center}
Dated: \today
\end{center}







\begin{abstract}
We report on multicolor excitation experiments with color centers in hexagonal boron nitride at cryogenic temperatures. We demonstrate controllable optical switching between bright and dark states of color centers emitting around 2eV. Resonant, or quasi-resonant excitation also pumps the color center, via a two-photon process, into a dark state, where it becomes trapped. Photoluminescence excitation spectroscopy reveals a defect dependent energy threshold for repumping the color center into the bright state of between 2.2 and 2.6eV. Photoionization and photocharging of the defect is the most plausible explanation for this behaviour, with the negative and neutral charge states of the boron vacancy potential candidates for the bright and dark states, respectively. Furthermore, a second zero phonon line, detuned by +0.4eV, is observed in absorption with orthogonal polarization to the emission, evidencing an additional energy level in the color center.
\end{abstract}

\vspace{5mm}
Color centers in wide bandgap semiconductors, most notably diamond and SiC, with atom like behavior and favorable spin properties, have long held potential for quantum photonic technologies, such as single photon sources and spin qubits \cite{Awschalom2018,Atature2018,doi:10.1002/qute.201900069}. More recently, color centres in hexagonal boron nitride (hBN) \cite{Tran2015,doi:10.1021/acsnano.6b03602,doi:10.1021/acs.nanolett.6b01987} have emerged as an alternative pathway for fundamental studies and the rapid progress reflects the wide interest and potential of this material system. These defects span a  wide spectral range, from the UV to the near-infrared \cite{doi:10.1021/acsnano.6b03602,PhysRevLett.117.097402,doi:10.1021/acs.nanolett.6b01368,doi:10.1021/acsnano.7b00638,PhysRevB.100.155419} and have attractive properties for quantum optics, including narrow linewidth, fast radiative recombination, stable emission and a relatively high fraction of photons emitted into the zero phonon line (ZPL). A particular benefit is the layered structure of hBN, which allows the fabrication of two-dimensional samples and integration of color centres directly with other 2D materials \cite{Geim2013,Liu2019} and/or integrated photonic platforms\cite{proscia2019scalable}. Furthermore, recent observations \cite{chejanovsky2019single,gottscholl2019room}  of an optically detected magnetic resonance suggests the potential to host spin qubits within a 2D material.

Typically, the choice of excitation laser plays a crucial role in the optical control of color centres. For example, in preparing and stabilising the charge state of color centres in diamond \cite{PhysRevLett.106.157601,PhysRevLett.109.097404,Aslam_2013,PhysRevLett.120.117401} and SiC\cite{Wolfowicz2017,Golter2017,PhysRevB.98.195202}.
In hBN there have been fewer studies in this area, but the choice of excitation energy has been shown to play a role in the absorption efficiency \cite{doi:10.1002/adma.201704237,Wigger_2019}, in stabilisation and enhancement of the PL emission \cite{Kianinia2018} and in photochemical modification of the defect \cite{doi:10.1021/acsphotonics.6b00736}.

In this work, we use multicolor excitation to investigate the photophysical dynamics of yellow emitting color centers in hBN. Typically green excitation is used to study such defects. The addition of a blue repump laser dramatically enhances the PL yield of certain color centres, but has a weak effect on others. We investigate two defects in detail using photoluminescence and photoluminescence excitation (PLE) spectroscopy. Through comparison with a rate equation model we find that the observed dynamics can be explained in terms of photo-induced switching between two metastable states. The color centers are pumped into a dark state by resonant, or quasi-resonant, excitation and require a higher energy laser to repump back to the bright state. The threshold for this repump is measured to be between 2.25 and 2.6eV, and is defect specific. We argue that the experimental observations are consistent with photoswitching between the dark neutral and bright negative charge states of a boron vacancy.

The sample consists of few-layer flakes of hBN drop-cast onto a Silicon substrate coated with 5nm of \ce{Al2O3}.\cite{doi:10.1021/acsnano.6b03602} The sample is annealed in nitrogen at 850\(\degree\)C for 8 minutes in a rapid thermal annealer. The optical experiments are performed at 5K in a closed-cycle cryostat using a micro-photoluminescence setup. The three lasers used in the experiment are co-aligned and coupled to a long-working distance microscope objective (N.A=0.8), which focuses the laser light to a diffraction limited spot of approximately 1\ce{\mu m} diameter. The luminescence from the sample is collected using the same objective and sent to a monochromator/CCD for spectroscopy, or filtered using a pair of tunable long and short pass filters for time-resolved and autocorrelation measurements. Blue and green excitation is provided by a 450nm (2.76eV) diode laser and a 532nm (2.33eV) diode pumped solid state (DPSS) laser, respectively. The DPSS laser can also be pulsed at variable repetition rate, with a pulse width of \(\sim\)50ps. For photoluminesence excitation spectroscopy (PLE) experiments a supecontinuum laser is filtered with an acousto-optic tunable filter to give a \(\sim\)2nm bandwidth and a pulsewidth of \(\sim\)5ps.

\begin{figure}
  \includegraphics[width=400pt]{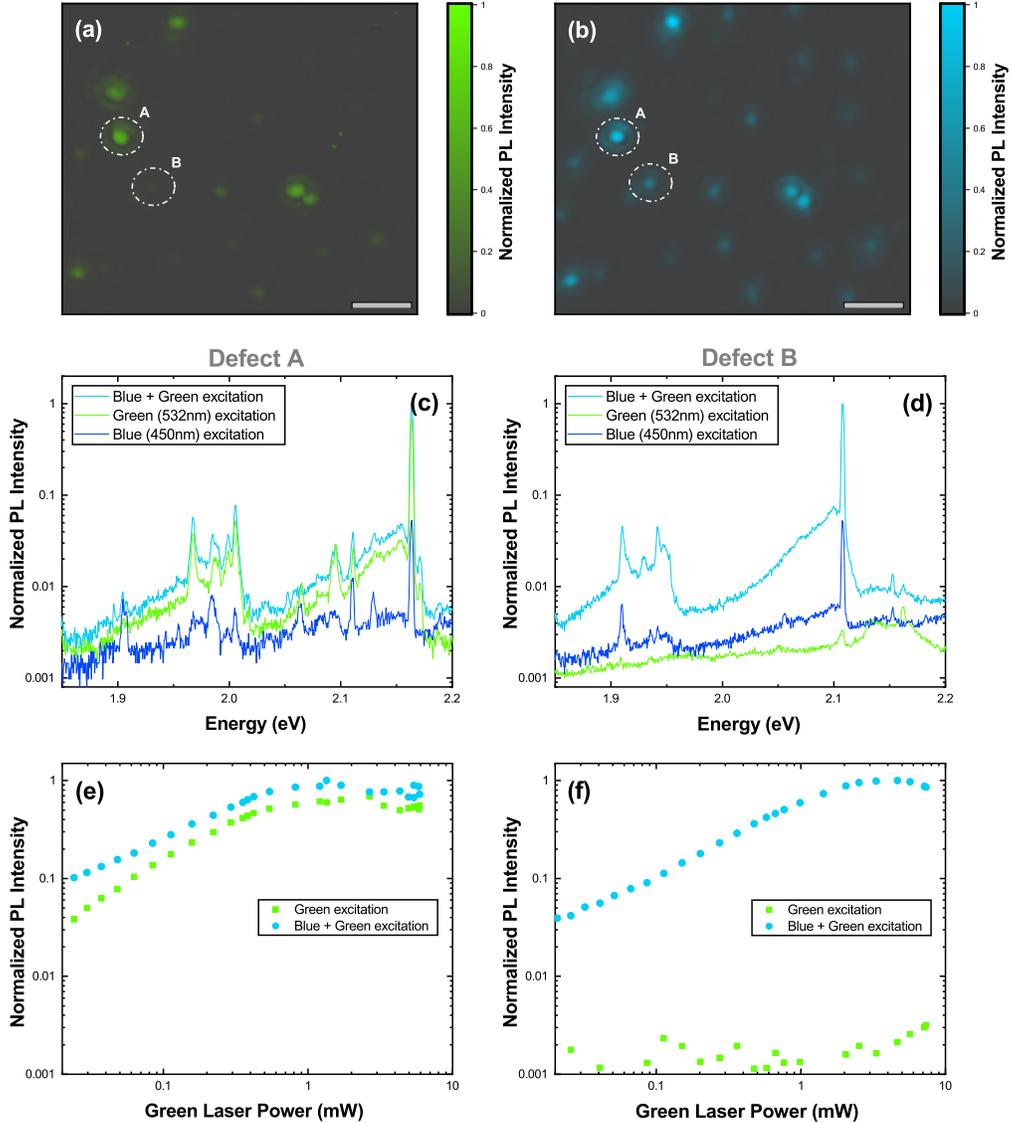}
  \caption{(a) and (b) Spatial maps of integrated photoluminescence emission (spectral range from 2.06 to 2.15eV) under (a) green only and (b) green plus blue excitation. The scale marker in (a) and (b) is 5\(\mu\)m. (c) and (d) Photoluminescence spectra recorded under green, blue and green plus blue, excitation conditions for the two defects circled in the PL maps, (c) defect-A and (d) defect-B. The green and blue excitation powers are  \(\sim700{\mu}W\). (e) and (f) Green laser power dependent intensity of the zero phonon line PL emission for (e) defect-A and (f) defect-B, with (blue circles) and without (green squares) the blue repump laser.}
  \label{fgr1}
\end{figure}

Figure~\ref{fgr1} illustrates the enhancement of the luminescence of color centers in hBN using multicolor excitation. Fig.~\ref{fgr1}(a) shows a typical PL map recorded with green excitation, which shows a number of bright spots resulting from individual, and small clusters of, color centers in hBN flakes. In Fig.~\ref{fgr1}(b), the same region of the sample is mapped using co-aligned green and blue continuous wave (CW) laser beams, resulting in an approximately two-fold increase in the number of luminescent centers. Two representative emitters are highlighted in the PL maps and defined as defect-A and defect-B. Defect-A appears bright in both maps, whereas defect-B is dark under green excitation, but bright when blue illumination is added. Further detail is shown in the PL spectra of the two defects under different illumination conditions in Fig.~\ref{fgr1}(c) and (d). Under green excitation the spectrum of defect-A [Fig.~\ref{fgr1}(c)] consists of a bright zero phonon (ZPL) line at 2.164eV and optical and acoustic phonon sidebands that are consistent with previous reports\cite{PhysRevB.100.125305,Wigger_2019,PhysRevB.99.020101}. Addition of the blue laser, results in a small enhancement of the overall PL intensity, only. With blue excitation alone, the ZPL intensity is reduced by a factor of \(\sim\)20, for the same excitation power. A comparison of the green laser power dependence of the ZPL of defect-A is made in Fig.~\ref{fgr1}(e), with and without the addition of the blue laser. The presence of the blue laser increases the PL intensity at all power of the green laser, including at powers beyond saturation, indicating that the blue laser is not just providing extra power, but increasing the PL yield of the defect.

In the case of defect-B, the choice of excitation conditions plays a much greater role. The PL spectrum, shown in Fig.~\ref{fgr1}(d), is similar to that of defect-A, with a single ZPL at ~2.107eV and similar optical and acoustic phonon sidebands. However, this is only the case for simultaneous excitation with blue and green lasers. With only blue excitation, the ZPL intensity is reduced by a factor of \(\sim\)20, whereas for only green excitation the intensity is reduced by a factor of \(\sim\)400. The power dependence [Fig.~\ref{fgr1}(f)] shows that even with a green laser power up to several mW the PL yield is still hundreds of times less intense without the blue laser. As illustrated by the PL maps, this behavior is not limited to a single defect, but is seen in approximately 50\% of defects we measure.

\begin{figure}
  \includegraphics{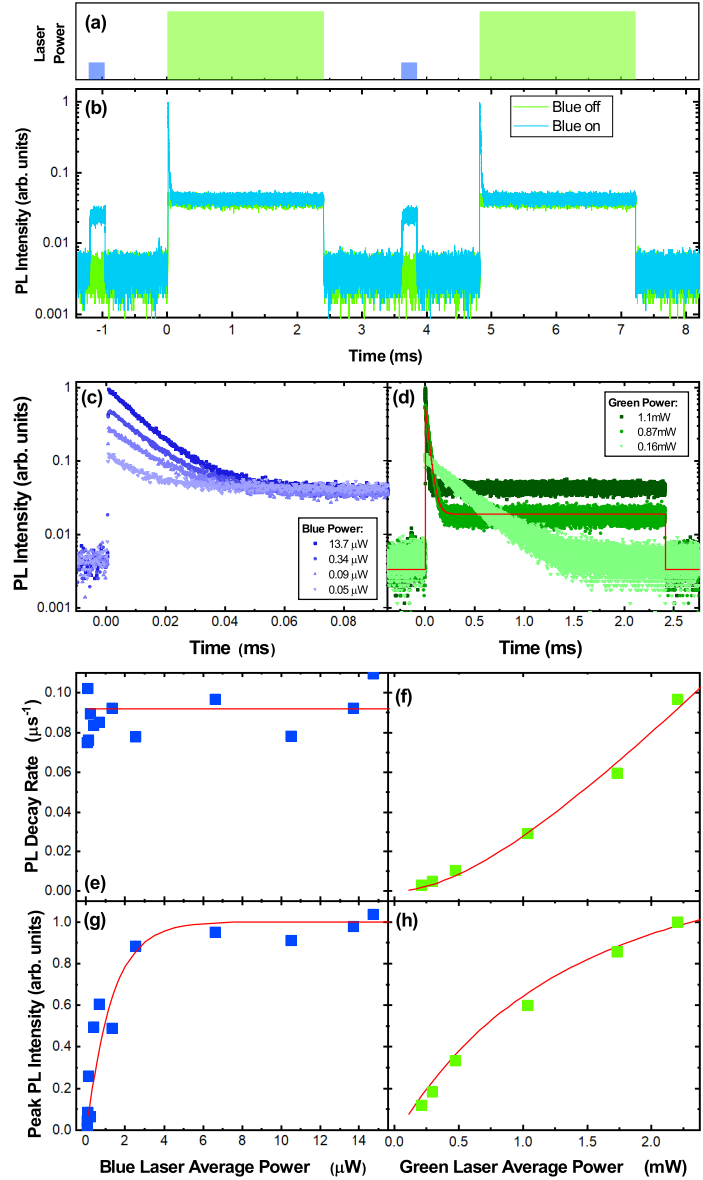}
  \caption{Time-resolved photoluminescence measurements of the zero-phonon line of defect-B. (a) Pulse-sequence used to excite photoluminescence. A blue pulse of width 0.5ms is followed by a green pulse of 2.5ms, with a repetition frequency of 208Hz. (b) Example PL intensity time trace. Without the blue pulse, the green laser excites low intensity emission. The blue pulse weakly excites PL, but also prepares the defect in the bright state. The subsequent arrival of a green pulse results in high intensity PL, which rapidly decays, indicating the defect is pumped into the dark state. (c) and (d) PL decay curves resulting from the onset of a green pulse for different (c) blue pump powers (fixed green power of 2.2mW) and (d) green pump powers (fixed blue power of 10.5\(\mu\)W). (e) and (f) Exponential decay rate as a function of (e) blue and (f) green pump power. (g) and (h) Peak ZPL intensity from the green laser pulse as a function of (g) blue and (h) green laser power. The red lines in (d) to (h) are fits to the data of the rate equation model.}
  \label{fgr2}
\end{figure}

To further investigate this behavior, we use time-resolved photoluminescence to study in detail the dynamics of defect-B. We first  investigate the role of blue and green excitation on the microsecond timescale. Fig.~\ref{fgr2}(a) illustrates the experiment, where a pulse train of alternating blue and green laser pulses are used to excite the defect, whilst the PL from the ZPL (filter bandwidth \(\sim\)1nm) is directed to an avalanche photon diode and photon-counting electronics. The pulse train consists of alternating 0.5ms and 2.5ms duration pulses of blue and green light with a repetition frequency of 208Hz. Fig.~\ref{fgr2}(b) plots a typical PL time trace recorded during this experiment, with and without the blue laser. With no blue laser pulses, the photoluminescence from the defect is weak and directly follows the intensity profile of the green laser. We term this the \textit{dark} state of the defect.

When the blue laser pulses are applied, weak PL is observed for the duration of the blue laser pulse. However, the blue pulse also prepares the defect in a \textit{bright} state, resulting in a strong PL signal at the start of the following green pulse. The initial peak PL increases linearly with low blue power, and saturates at a few \ce{\mu W} [Fig.\ref{fgr2}(g)]. Under green illumination, the bright state lives for tens of $\mu$s. The PL decays exponentially with a decay-rate that is quadratic with green laser power, indicating that the {\it bright} to {\it dark} transition is mediated by two-photon process [Fig.\ref{fgr2}(f)].

\begin{figure}
  \includegraphics[width=\textwidth]{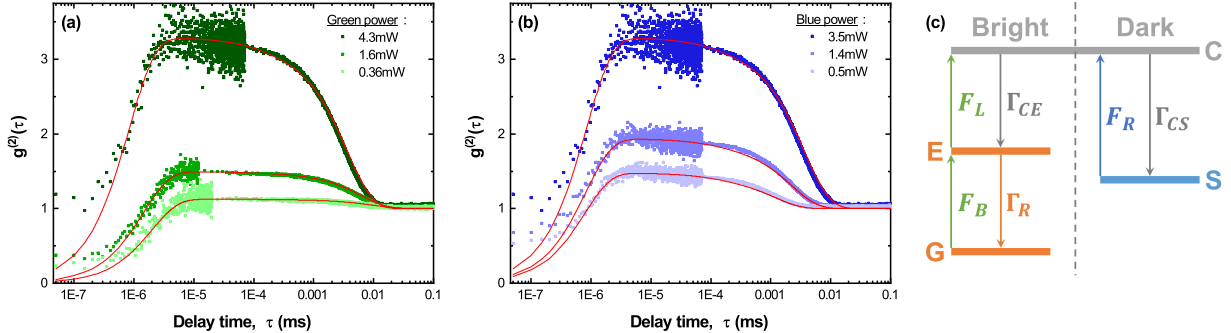}
  \caption{Second order autocorrelation function \(g^2(\tau)\) of defect-B with (a) fixed blue power of 0.5mW and varying green power and (b) fixed green power of 4.3mW and varying blue power. The red lines show fits to the data using the model outlined in the text. (c) Rate equation model of photoionization and recharging.}
  \label{fgr3}
\end{figure}

For qualitative understanding, we construct a simple rate equation model motivated by the experimental observations and illustrated in Fig.~\ref{fgr3}(c). The model consists of four energy levels: \(G\) and \(E\) are the ground and excited levels of the bright state; \(S\) represents the dark shelving state of the defect; and \(C\) represents the conduction band or some higher energy levels of the system, as discussed in further detail below. \(\Gamma_R\) is the radiative decay of the bright state and \(\Gamma_{CS}\) and \(\Gamma_{CE}\) are non-radiative transitions which link the dark and bright states. For each transition there is a corresponding optical pump rate: \(F_B\) is the bright state pump rate, \(F_L\) is the rate at which electrons are pumped out of the bright excited state and \(F_R\) is the rate at which the color center is pumped out of the shelving state. The resulting system of four differential equations is solved numerically to calculate the photoluminescence, \({\propto}E\Gamma_R\) (See Supporting Information for equations and further model details). The model is used to fit the data in Fig.~\ref{fgr2} (e) to (h) and shows good agreement with the main experimental observations. Firstly, the rate of the dark to bright process, plotted in Fig.~\ref{fgr2}(f), is ~quadratic with power, in agreement with a two-photon bright-to-dark process. Secondly, as shown in Fig.~\ref{fgr2}(g) a blue pulse prepares the defect in the bright state, with a probability proportional to power, indicating a one photon process that requires an energy greater than 2.33eV (green), but less than 2.76eV (blue).

To further verify this model, measurements are made of the second order auto-correlation, \(g^{(2)}(\tau)\), of defect-B for different blue and green excitation powers. In Fig.~\ref{fgr3}(a) \(g^{(2)}(\tau)\) is plotted for fixed blue power of 0.36mW and three different green powers. For comparable green and blue powers \(g^{(2)}(\tau)\) is close to that expected for a two-level emitter. However, increasing the green power leads to pronounced bunching with an exponential decay on the order of microseconds. Similar bunching behavior has been observed previously for color centers in both hBN \cite{PhysRevB.96.121202,doi:10.1021/acsnano.6b03602} and other wide band semiconductors \cite{PhysRevA.81.043813,Neu:12} and attributed to the presence of metastable dark states. In Fig.~\ref{fgr3}(b) \(g^{(2)}(\tau)\) is plotted for high fixed green power of 4.3mW and three different blue powers. When the blue power is low, the bunching is strong, but increasing the blue power counteracts the tendency of the green laser to pump the defect into the dark state and \(g^{(2)}(\tau)\) close to the 2-level case is recovered. Again, similar observations have been made for the repumping of a color center in hBN, but with different emission and pump energies \cite{Kianinia2018}. The \(g^{(2)}(\tau)\) function is calculated from the rate equation model with the Quantum Toolkit in Python package \cite{JOHANSSON20121760,JOHANSSON20131234} (QuTIP) using the built-in function qutip.coherence{\textunderscore}function{\textunderscore}g2. This applies a master equation method to calculate \(g^2(\tau)\) from a Hamiltonian, list of collapse operators for incoherent transitions and timescale. As the numerical values of the energy levels are unimportant for this calculation, and none of the transitions are being driven coherently and close to resonance, a trivial Hamiltonian, \(\hat{H} = \hat{O}\) is used and the incoherent optical pumping operators are included with the collapse operators (See Supporting Information for example code). By keeping the radiative and non-radiative transition rates constant and only varying the pump rates in proportion to the experimental powers, very good fits to the data could be found, as plotted in Fig.~\ref{fgr3}(a) and (b) (Note: the same model parameters are used to fit both the time-resolved PL and \(g^{(2)}(\tau)\) data).

\begin{figure}
  \includegraphics[width=\textwidth]{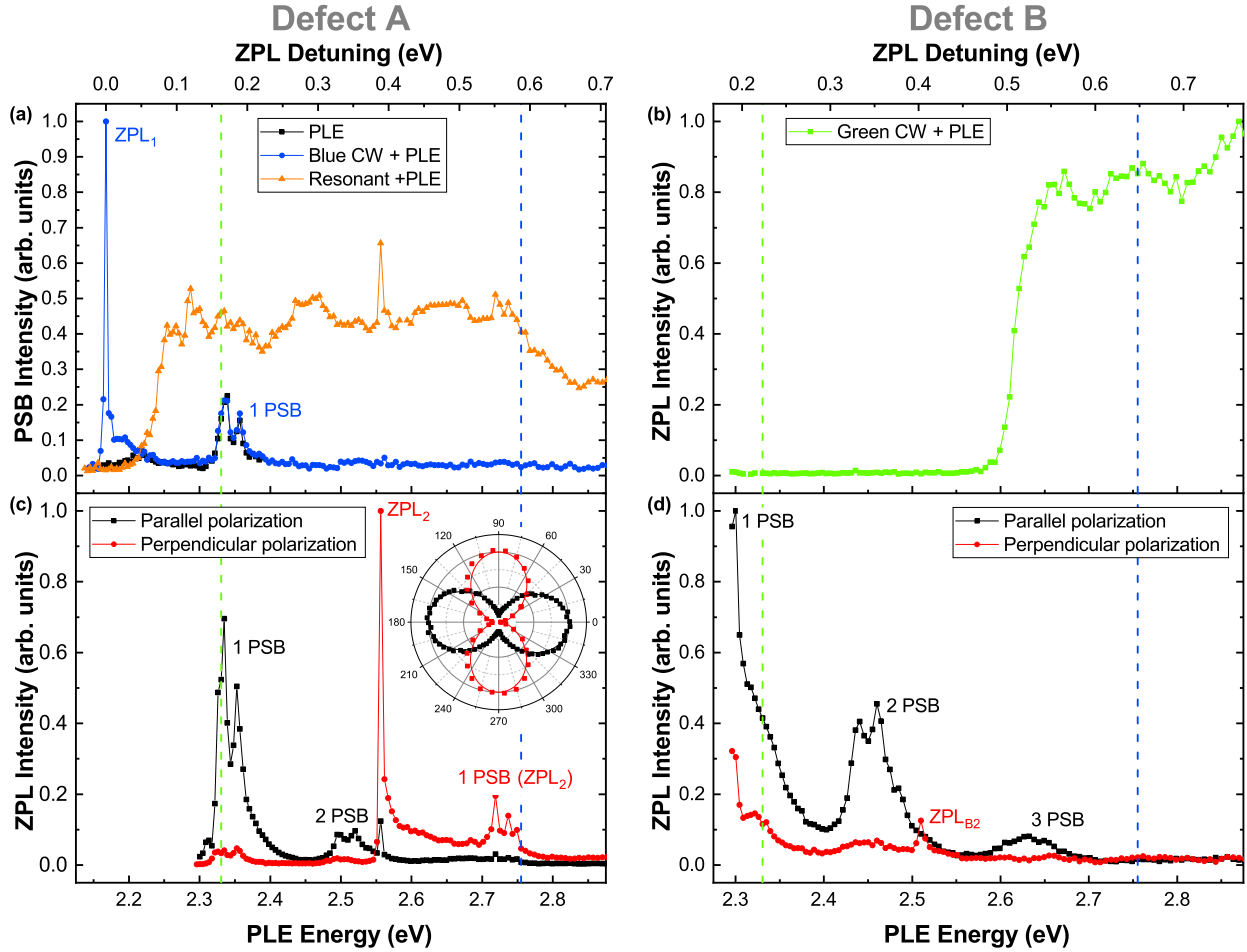}
  \caption{Photoluminescence Excitation (PLE) Spectroscopy.  (a) PLE spectrum of defect-A with (blue circles) and without (black squares) blue CW laser and with fixed resonant excitation (orange triangles).  (b) PLE spectrum of defect-B with fixed green CW laser (green squares) (c) PLE spectrum of defect-A for laser polarization parallel (black squares) and perpendicular (red circles) to the emission from \(ZPL_1\). The inset shows a polar plot of intensity of \(ZPL_1\) as a function of the excitation polarization angle when resonant with \(ZPL_1\) (573nm, black squares) and \(ZPL_2\) (485nm, red circles). The black and red solid lines show a \(\sin^2 \theta \) fit to the data. (d) PLE spectrum of defect-B for laser polarization parallel (black squares) and perpendicular (red circles) to the emission from \(ZPL_1\).}
  \label{fgr4}
\end{figure}

From the two-color experiments it is clear that the photon energy of the excitation laser determines if the color center is in a bright or dark state. We therefore perform PLE spectroscopy with a tunable pulsed supercontinuum laser, to investigate how the optical pumping depends on photon energy. We first measure the energy threshold of the repump process for defect-B, plotted in Fig.~\ref{fgr4}(b). The defect is excited by both the green CW laser and the tunable laser and reveals a clear energy threshold for the repump process of \(\sim\)2.6eV. To understand why defect-A is bright under green only illumination, a similar measurement is made. The PLE spectrum is measured with and without the addition of the blue CW laser, as plotted in Fig.~\ref{fgr4}(a), whilst recording the intensity of the phonon sideband (PSB) emission. Without the blue laser it is not possible to resonantly excite the defect. The threshold is measured as 2.25eV by exciting the defect resonantly, whilst sweeping the repump laser energy, as shown in Fig.~\ref{fgr4}(a). Although the emission energy of the ZPL of defect-A and B differs by only \(\sim\)50meV, the difference in the repump energy threshold is \(\sim\)350meV. Nevertheless, the PLE measurements suggest that the two defects can be described by the same model, but with different energy dependence of the repump rate, \(F_R\).

The simple model describes the data well and we propose the following interpretation in terms of charge state conversion. The green laser efficiently excites the defect from the ground (\(G\)) to excited state (\(E\)). A second green photon can further excite the electron into the conduction band, \(C\). From \(C\), the electron can be recaptured recharging the color center into the bright state; or the electron is lost or trapped by another defect state, shelving the color center in a dark charge state,  \(S\). In this picture, the bright state is negatively charged with respect to the dark state. A similar process involving the valence band is also compatible with the data. A similar two-photon ionization process is responsible for photo-switching from the negative to neutral charge state of NV-centers in diamond  \cite{PhysRevLett.106.157601,PhysRevLett.109.097404,Aslam_2013} and out of the neutral state of the silicon divacancy in SiC \cite{Golter2017,Wolfowicz2017}. In the case of NV centers, a two-photon process is also responsible for repumping. However, in our measurements we find that the repump process is one-photon assuming the photon energy is sufficiently large to overcome the photo-charging threshold of, 2.25eV and 2.6eV for defects A and B, respectively. The plateau-like dependence of the repumping process suggests the involvement of either the conduction or valence band, further supporting this photocharging picture  \cite{Golter2017,PhysRevB.98.195202}.

The PLE measurements reveal further information about the nature of the color centers in hBN. In Fig.~\ref{fgr4}(c), the PLE spectrum of defect-A is measured, for excitation polarization parallel and perpendicular to the ZPL emission. When the laser is polarized parallel to the ZPL emission, absorption resonances are observed in two bands detuned by \(\sim\)170meV and \(\sim\)340meV from the ZPL, which can be attributed to one and two optical phonon-assisted absorption, respectively. With orthogonal polarization the PLE spectrum is very different, with only weak absorption into the PSB, but with a strong and spectrally narrow absorption resonance  detuned from \(ZPL_1\) emission by 400meV, evidencing a higher energy level in the electronic structure of the color center, labelled \(ZPL_2\) in Fig.~\ref{fgr4}(c). The PLE spectrum is  similar to that of \(ZPL_1\), including a PSB detuned by \(\sim\)170meV. Furthermore, the polar plot in the inset of Fig.~\ref{fgr4}(a) shows that the two ZPLs are orthogonally polarised, with an \(89\degree\) misalignment of the absorption dipoles of \(ZPL_1\) and \(ZPL_2\) extracted from \(\sin^{2}\theta\) fits. Although the absorption resonance is clearly observed when detecting at the energy of \(ZPL_1\), there is no sign of PL emission at the energy of \(ZPL_2\) when exciting with the blue laser (see Supporting Information Fig. S1). These observations explain previous measurements where depending on the detuning of the laser energy from the ZPL the absorption and emission dipoles can be aligned or misaligned by up to 90\(\degree\)\cite{doi:10.1021/acs.nanolett.6b01987,PhysRevLett.119.057401}. As shown in Fig. S1 of the Supporting Information, when exciting with energy above(below) \(ZPL_2\) the absorption polarization is perpendicular (parallel) to the emission of \(ZPL_1\).

A similar PLE measurement of defect-B is presented in Fig.~\ref{fgr4}(d), where a low power blue excitation is used to maintain the color center in its bright state. As with defect-A, when the polarization of the tunable laser is aligned parallel to the ZPL, absorption peaks corresponding to phonon assisted processes are observed. In this case, the relative efficiency of the two phonon sideband is greater than for defect-A and the three phonon sideband is also observed. For the orthogonal polarization the absorption is considerably less efficient at all energies, but there is a weak absorption peak at \(\sim\)2.51eV, labelled \(ZPL_{B2}\) in Fig.~\ref{fgr4}(d). As with defect-A, the detuning of this absorption peak is 403meV, suggesting a similar origin. Similar measurements of other defects are presented in the Supporting Information. From a survey of 16 individual defects, 11 show a similar absorption peak in the PLE spectra with orthogonal polarization to the ZPL emission, with an average energy detuning of 0.5eV and standard deviation of 0.1eV.

Measurements of color centers in diamond again provide some clues as to the origin of \(ZPL_2\). In the photocharging cycle of neutral and negatively charged NV centres, resonant excitation both at the ZPL energy of \(NV^0\) and at a higher energy level of the \(NV^-\), results in an increase in the PL yield from \(NV^-\) \cite{PhysRevLett.109.097404}. In negatively charged silicon vacancy centers a higher energy level with perpendicular polarization was identified, but was not observed in PL due to selection rules of the energy levels \cite{PhysRevB.89.235101}. Further work is required to confirm the origin of the transition reported here, but it provides key extra information to compare with theoretical calculations.

Regarding the possible identity of the color centers here, we compare the observations to Density Functional Theory (DFT) calculations in the literature. The defects have two in-plane polarized ZPL transitions with orthogonal polarization. Whilst several defect species (e.g. Boron dangling bond\cite{PhysRevLett.123.127401}) have transitions similar to 2 eV, most are out-of-plane polarized and have an orbital degeneracy of one. A possible candidate is the $^{3}A_{2}^{'}$ $\longleftrightarrow$ ${}^{3}E'$ transition of the negatively charged boron vacancy, $V_B^-$. The upper energy level has two orbital states, and a symmetry breaking would result in two orthogonal linearly-polarized transitions at energy of 2.1-2.3eV \cite{ivdy2019ab,doi:10.1021/acsphotonics.7b01442}. We could not find any calculations of the expected splitting, that we measure to be $\sim$0.4eV, which seems large. Furthermore, the neutral boron vacancy $V_B^0$ is predicted to have no optically active transitions below 3.5eV \cite{doi:10.1021/acs.nanolett.7b04819,PhysRevB.83.144115}, and the charge transition level for the neutral to negative state of the boron vacancy has been calculated (with reference to the valance band maxima) 1.5eV\cite{PhysRevB.97.214104}, and 2.1eV\cite{Strand_2019} for bulk hBN, and 2.4eV \cite{doi:10.1021/acsphotonics.7b01442} for a monolayer. This is in agreement with our model of photoionization and recharging. With sufficient energy a photon from the repump laser can convert the dark \(V_B^0\) into the bright $V_B^-$ by promoting an electron from the valence band, whereas two photons are required to remove an electron to the conduction band and convert \(V_B^-\) to \(V_B^0\). The \(\sim\)350meV variation in the charging threshold could result from local variations, for example, from strain, the number of layers of the flake that hosts the defect, \cite{PhysRevMaterials.2.124002} or interaction with the substrate \cite{PhysRevMaterials.3.083803}. However, we note that the repumping threshold could also represent the charging threshold of a nearby trap state, and many defect species have charging thresholds in this region \cite{Strand_2019,PhysRevB.97.214104}.

In conclusion, illuminating a color-center in hBN on, or near, resonance drives the color center into a dark state. Applying a second laser with a photon energy exceeding a defect specific threshold repumps the color center back to the bright state. Phenomenologically, the dynamics can be understood in terms of photoswitching between two different charge states of the same defect, as is the case for color centers in diamond and SiC\cite{Golter2017,Wolfowicz2017,PhysRevB.98.195202}. Furthermore, PLE spectroscopy reveals a sharp absorption resonance, at approximately 400meV higher energy than the ZPL. The ZPL and charge transition threshold energies are consistent with calculated values for the boron vacancy \cite{ivdy2019ab,doi:10.1021/acsphotonics.6b00736,Strand_2019}. These results are of practical use for enhancing the PL yields in both future devices and experimental studies, particularly under resonant excitation\cite{Konthasinghe:19} and can play an important role in the identification of color centers in hBN.

\begin{acknowledgement}

This work was supported by the Engineering and Physical Sciences Research Council [Grant numbers EP/S001557/1 and EP/026656/1].

\end{acknowledgement}

\begin{suppinfo}
Additional experimental measurements on defect-A, PLE measurements of $ZPL_2$ in other defects, further model details and example QuTIP notebook.
\end{suppinfo}

\bibliography{hBN_blue-green}

\end{document}